\documentclass[twocolumn,twoside,slac]{revtex4}
\usepackage{graphicx}
\usepackage{fancyhdr}
\begin{document}

%Title of paper
\title{Comments on Likelihood fits with variable resolution}

\author{Giovanni Punzi}
\affiliation{ Scuola Normale Superiore and INFN, 56100 Pisa, Italy}

\begin{abstract}
Unbinned likelihood fits are frequent in Physics, and often involve complex functions with several components. We discuss the potential pitfalls of situations where the templates used in the fit are not fixed but depend on the event observables, as it happens when the resolution of the measurement is event--dependent, and the procedure to avoid them. 
\end{abstract}

\maketitle
\thispagestyle{fancy}

When several categories of events are present in the same data sample, an unbinned Maximum Likelihood fit is often used to determine the proportion and the properties of each class of events. This procedure makes use of ``templates", representing the probability distribution of the observables used in the fit for each class of events.  In the simplest cases the templates are completely determined by the values assigned to the parameters of the fit, but frequently a more sophisticated approach is chosen where templates vary on an event by event basis, according to the resolution of the measurement for that particular event. These variations are due to the dependence of resolution on extra variables, that change on an event-by-event basis . This may happen, for instance, when events are recorded by a detector that has different resolutions in different regions within its acceptance.  

A common example of this kind of fit in HEP is given by lifetime and/or mass fits (see \cite{lifemass_papers} for a sample list of recent experimental papers), where variations in resolution occur as a consequence of different configuration of each individual decay. The same kind of issue hovewer is likely to arise in other situations.

The purpose of this short paper is to point out some potential pitfalls
in this kind of fitting procedure. I will illustrate the point with reference to a simple toy problem.

\section{A toy problem}

Consider an experiment in which two types of events, A and B, can occur. Let $f$ be the fraction of type--A events, that is, the probability of a generic event to be of type A. We want to extract a measurement of $f$ from a given sample of data. 
In order to do this, we measure the value of an observable $x$, having the following
probability distributions:
\begin{eqnarray*}
p(x|A) = N(0, \sigma)\\
p(x|B) = N(1, \sigma)
\end{eqnarray*}
Where $\sigma$ is a known constant and $N(\mu,\sigma)$ is the normal distribution

 This problem is easily solved using an ``unbinned Likelihood fit". This consists of maximizing the Likelihood function:
 \begin{equation}\label{eq:fixed_L}
 L( f ) = \prod_i{( f  N(x_i,0, \sigma) + (1- f ) N(x_i,1, \sigma) )}
 \end{equation}
 with respect to the required parameter $f$ (here $N(x,\mu,\sigma)$ indicates the gaussian function in the variable $x$). This is very simple to perform with the help of a numerical maximization program. 

\begin{figure}[h]
\includegraphics[width=2in]{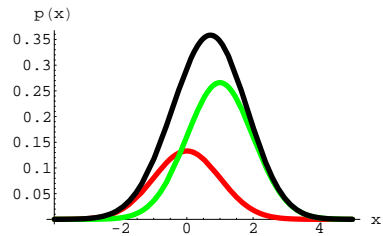}%
\caption{\label{fig:A-B} Probability distribution of $x$ for the toy problem described in the text. Contribution of type--A and type--B events are also shown.}
\end{figure}

\begin{figure}[h]
\includegraphics[width=2in]{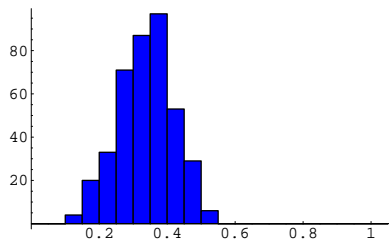}%
\caption{\label{fig:MC_fixed} Distribution of ML estimate of the fraction $f$ of type-A events (see text)}
\end{figure}

Let's make a specific numeric example, where $f=1/3$ and $\sigma=1$ (see illustration in Fig.~\ref{fig:A-B}), and the size of the data sample is 150 events. By repeteadly generating MC  samples of 150 events each, we obtain the distribution of the Maximum Likelihood estimator of $f$, which is shown in Fig.~\ref{fig:MC_fixed}.  

Its mean is $0.3368\pm 0.0041$ and $SD=0.083$, in agreement with expectations of $0.3333$ and $0.088$ respectively (the latter coming from Fisher information ).

\section{A toy problem, with variable resolution}

\begin{figure}[h]
\includegraphics[width=2in]{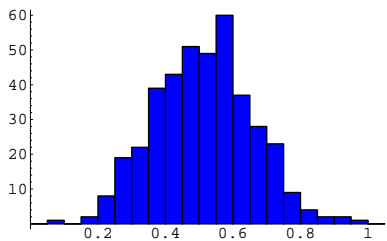}%
\caption{\label{fig:MC_variable} Distribution of ML estimate of the fraction $f$ of type-A events, obtained from a "conditional Likelihood"}
\end{figure}

Let's now suppose that the resolution of $x$ is not constant, but rather depends on the event: we are assuming that each event $x_i$ comes together with an individual value of $\sigma$ (let it be $\sigma _i$). This situation is encountered in many real--life problems, and the common approach found in the literature is to simply modify the Likelihood function as follows:
\begin{equation}\label{eq:cond_L}
L(f) = \prod_i{f N(x_i,0, \sigma_i) + (1- f ) N(x_i,1, \sigma_i))} 
\end{equation}
This looks like a pretty obvious generalization of expression (\ref{eq:fixed_L}). To test it in our toy problem, we modified our toy MC from previous example, by making $\sigma$ fluctuate at each event within an arbitrarily chosen range ($1.0$ to $3.0$), and again made repeated simulated experiments of 150 events each, maximizing the Likelihood expression (\ref{eq:cond_L}) to estimate $f$.  The result of this test is shown in Fig.~\ref{fig:MC_variable}, and rather surprisingly, shows a very large bias with respect to the true value of $f$.

This may seem really odd, until one realizes that this new problem is very different from the previous one. 
Our problem now has actually two observables: each observation consists of the pair of values $(x_i,\sigma_i)$ rather than just $x_i$, and its probability density depends on both. This means that the Likelihood must now be written based on the probability distributions of the $(x_i,\sigma_i)$ {\em pair}:
\begin{equation}\label{eq:generic_L}
 L(f ) = \prod_i {f p(x_i,\sigma_i |A)+(1-f) p(x_i,\sigma_i |B)}
\end{equation}
Remembering that $p(x_i,\sigma_i |X)= p(x_i|\sigma_i ,X)p(\sigma_i|X)$ we can  write the correct expression of the Likelihood for our problem as:
\begin{eqnarray}\label{eq:full_L}
 \nonumber L(f) = \prod_i {f N(x_i,0, \sigma_i)p(\sigma_i |A)}\\
 \mbox{ } +(1-f) N(x_i,1,\sigma_i)p(\sigma_i |B)
\end{eqnarray}
where $p(\sigma_i|X)$ is the {\em pdf} of $\sigma_i$ for events of type $X$, an element that was absent in eq.~(\ref{eq:cond_L}); in fact, comparing the two expressions shows that~(\ref{eq:cond_L}) is actually the {\em conditional} probability distribution $p(x_i | \sigma_i,f)$ (one might call it ``conditional Likelihood") rather than the full distribution $p(x_i ,\sigma_i | f)$. 
The difference matters for fitting unless it happens that the distribution of $\sigma_i$ is the same for all types of events: $p(\sigma_i |A)= p(\sigma_i |B)$. In that case, $p(\sigma_i)$ can be factored out, and the incomplete Likelihood of eq.~(\ref{eq:cond_L}) differs from the true Likelihood just by a factor independent of $f$, that does not affect the maximization.

In the specific MC test reported above, we simulated a resolution 1.5 times worse for events of type B than for type A, setting the $\sigma_i$ distribution as flat between 1 and 2 for A-type events, and flat between 1.5 and 3 for type-B events. We intentionally avoided saying this explicitly before, in order to put the reader in the typical situation encountered in reality, where no attention is payed to the distribution of those resolutions for the different classes of events considered in the fit.  It turns out from our example that this may lead to very biased results.

In summary, expression~(\ref{eq:cond_L}) simply does not work for fitting, and by a large amount: it can be said to belong to that particular class of solutions nicely defined in \cite{Heinrich} as `SNW solutions'. 

\begin{figure}[h]
\includegraphics[width=2in]{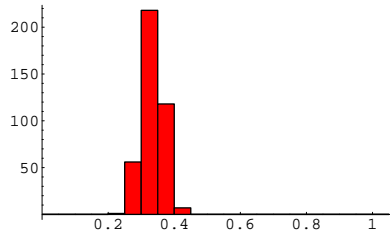}%
\caption{\label{fig:MC_full} Distribution of ML estimate of the fraction $f$ of type-A events, using the full Likelihood function}
\end{figure}

Conversely, if we use in fitting the correct expression of the Likelihood (eq.~\ref{eq:full_L}) we get the result shown in fig.~\ref{fig:MC_full}, showing a negligible bias. The resolution of the fit is also much better, as the difference in the distributions of the $\sigma$ themselves gets exploited in separating the two samples; this however is a minor point in comparison with the bias issue.

\section{Additional tests}\label{sec:more_tests}

One may wonder at what features of the distributions make for a large bias. Table~\ref{tab:varied} shows results for a few variants of the original problem. 
Tests include:
\begin{itemize}
\item{Equal-width ranges of $\sigma$.}
\item{Disjoint $\sigma$ ranges.}
\item{Constant, but different $\sigma$ for A and B.}
\item{Constant, and close $\sigma$'s for A and B.}
\item{Same-mean $\sigma$ distribution with different widths.}
\item{Only one type of events has variable sigma.}
\end{itemize}

\begin{table}% add [H] placement to break table across pages
\begin{centering}
\caption{\label{tab:varied} Results of MC fitting tests.}
\footnotesize
 \begin{tabular}{|cc||cc|cc|}
 \hline
  \multicolumn{2}{|c||}{Resolutions}&  \multicolumn{2}{|c|}{``conditional" L (\ref{eq:cond_L})} & \multicolumn{2}{|c|}{True Likelihood} \\ 
$\sigma_A$ & $\sigma_B$& $\tilde{f}_A$ & $\sigma(\tilde{f}_A)$ & $\hat{f}_A$ & $\sigma(\hat{f}_A)$\\
\hline
$1.0$ & $1.0$ & $0.336\pm0.003$& 0.08 & &\\ 
$\left[1.0,2.0\right]$ & $\left[1.5,3.0\right]$& $0.514\pm0.007$& 0.14 & $0.335\pm0.002$& 0.03\\ 
$\left[1.0,2.0\right]$ & $\left[1.5,2.5\right]$& $0.474\pm0.007$& 0.14 & $0.335\pm0.002$& 0.03\\
$\left[1.0,2.0\right]$ & $\left[2.0,3.0\right]$& $0.579\pm0.008$& 0.15 & $0.333\pm0.000$& 0.00\\
$ 1.0 $ & $ 2.0 $& $0.645\pm0.006$& 0.12 & $0.333\pm0.000$& 0.00\\
 1.0 & 1.1 & $0.374\pm0.004$& 0.08 & $0.333\pm0.000$& 0.00\\
$\left[0.5,3.5\right]$ & $\left[1.5,2.5\right]$& $0.330\pm0.006$& 0.12 & $0.332\pm0.002$ & 0.03\\ %test10
$ 1.0 $ & $ \left[1.0,2.0\right]$& $0.482\pm0.009$& 0.09 & $0.333\pm0.000$& 0.00\\ %test8
\hline\hline
  \multicolumn{2}{|c||}{($\sigma_A$ actually $=1.$) }&  \multicolumn{2}{|c|}{modified L (\ref{eq:cond_L_fixed})} & \multicolumn{2}{|c|}{True Likelihood} \\ \hline 
$ 1.0 $ & $ \left[1.0,2.0\right]$& $0.374\pm0.004$& 0.08 & $0.333\pm0.000$& 0.00\\ %test8
$\left[0.5,3.5\right]$ & $\left[1.0,2.0\right]$& $0.414\pm0.004$& 0.08 & $0.332\pm0.003$& 0.03\\ %test9

\hline
\end{tabular}
 \end{centering}
 \end{table}

In almost every tried situation we found expression~(\ref{eq:cond_L}) to return largely biased results. The exception occurs when the average $\sigma$ is the same; the resolution on $f$ is however much worse than with the correct expression. It looks like the most important element is the difference between the average values of $\sigma$ for the different samples; the actual variability within each sample seems less important. 

A simpler situation exists, that is pretty common in practice, where one has just one signal component over a background, and the signal distribution contains a variable sigma, while the background is represented just by a fixed template. In this case, expression~(\ref{eq:cond_L}) becomes:
\begin{equation}\label{eq:cond_L_fixed}
L(f) = \prod_i{f N(x_i,0, 1) + (1- f ) N(x_i,1, \sigma_i))} 
\end{equation}
This expression of L is of course still incorrect, but it better describes reality at least for one of the two event categories by incorporating explicitly the information that it has a fixed sigma. Here a variable template appears just in one component, and being this the simplest configuration with a variable template, it is interesting to ask whether it yields a reasonable approximation of the correct results.

If we apply this new Likelihood expression to the last tested case, ($\sigma_A=1.0$ and $\sigma_B \in \left[1.0,2.0\right]$), we find that the result is still biased, although to a lesser extent (Tab. \ref{tab:varied}). This shows that  the distribution of $\sigma$ must be kept into account even in the simplest situation, where it appears in only one component of the fit.

The mechanism underlying this problem is easier to see by looking at a variant of the previous case. Suppose that resolutions are the same as above, but for events of type-A the variable $\sigma_i$ is distributed over a wide range (0.5-3-5);  this is not the actual value of the resolution for those events, that is still fixed at 1, so for type-A events it represents just an additional meaningless number. This is a definite possibility in a real case, where the nature of type--A events may be so different from type--B to produce meaningless values for the resolution estimator $\sigma_i$, that was designed to work for type--B events -- remember that the distribution of A is given as fixed.
Note that the expression used (\ref{eq:cond_L_fixed}) does know that much, and correctly disregards the value of $\sigma_i$  in the A hypothesis. For events of type B, the variable $\sigma_i$ correctly represents the sigma, event by event, of the observable $x$, and the L function correctly accounts for this, too. 

It may come as a surprise that the result is largely biased.  
The reason for this rather spectacular failure is that the second piece of L, related to B-type events, gets confused by the presence of the events of type--A with meaningless values of sigma: they unavoidably enter both terms of L during the calculation. The conclusion is: whenever you include $\sigma_i$ in your Likelihood expression, even for just {\em one} class of events, you must also account for its distribution, and you must do so for {\em all} event classes.

\section{Conclusions}

Whenever the templates used in a multi-component fit depend on additional observables, one should always use the correct, complete Likelihood expression (\ref{eq:full_L}), including the explicit distributions of all observables for all classes of events. This is necessary even if just one of the components is based on a variable $\sigma$.  The simpler expressions that are commonly used should be considered unreliable unless one can show that the distribution of the variable $\sigma$ is the same for all components.

A more general consideration suggested by the examples discussed above is that one should always be wary of ``intuitive" modifications of a Likelihood function. For every given problem there is only one correct expression for the Likelihood (up to a multiplicative constant factor), and it is crucial to verify in every case that the expression used is the right one, rather than rely on intuition.

\end{document}